\documentclass[twocolumn,pre,superscriptaddress]{revtex4}

\usepackage{dcolumn}
\usepackage{amsmath}

\usepackage{graphicx}

\renewcommand{\d}{\mathrm{d}}
\newcommand{\Ord}{\mathrm{O}}
\newcommand{\e}{\mathrm{e}}
\newcommand{\half}{\mbox{$\frac12$}}

\newcommand{\av}[1]{\langle#1\rangle}

\newcommand{\defn}{\textit}

\begin{document}

\title{Bicomponents and the robustness of networks to failure}
\author{M. E. J. Newman}
\affiliation{Department of Physics, University of Michigan, Ann Arbor, MI
  48109}
\affiliation{Center for the Study of Complex Systems, University of
  Michigan, Ann Arbor, MI 48109}
\author{Gourab Ghoshal}
\affiliation{Department of Physics, University of Michigan, Ann Arbor, MI
  48109}
\affiliation{Michigan Center for Theoretical Physics, University of
Michigan, Ann Arbor, MI, 48109}
\begin{abstract}
  A common definition of a robust connection between two nodes in a network
  such as a communication network is that there should be at least two
  independent paths connecting them, so that the failure of no single node
  in the network causes them to become disconnected.  This definition leads
  us naturally to consider bicomponents, subnetworks in which every node
  has a robust connection of this kind to every other.  Here we study
  bicomponents in both real and model networks using a combination of exact
  analytic techniques and numerical methods.  We show that standard network
  models predict there to be essentially no small bicomponents in most
  networks, but there may be a giant bicomponent, whose presence coincides
  with the presence of the ordinary giant component, and we find that real
  networks seem by and large to follow this pattern, although there are
  some interesting exceptions.  We study the size of the giant bicomponent
  as nodes in the network fail, using a specially developed computer
  algorithm based on data trees, and find in some cases that our networks
  are quite robust to failure, with large bicomponents persisting until
  almost all vertices have been removed.
\end{abstract}
\maketitle

The robustness of connections in networks has been studied extensively in
the physics community, in part because of its practical importance in
settings such as communications networks and
epidemiology~\cite{AJB00,CEBH00,CNSW00}.  The typical approach is to
consider the largest set of vertices in a network that are connected to one
another by at least one path, the so-called \defn{giant component}, and
examine how its size varies as vertices are removed from the network.  This
can be thought of as a simple model for the performance of, for instance, a
communication network such as the Internet under failure of vertices.  The
vertices removed can be chosen at random, as if failing because of random
technical faults, or in a targeted fashion, as if an adversary were
deliberately removing them in an effort to destroy network connectivity.

In real-world situations, however, it is often considered inadequate to
rely for communication on just a single path between vertices.  Most large
organizations, for example, connect to the Internet using a strategy called
\defn{multihoming}, in which they maintain two or more independent data
connections to the network so that the failure of any one connection will
not leave them disconnected.  More generally, the connection between two
vertices is considered robust if there are at least two independent paths
between them, so that the failure of no single vertex in the network can
cause the two to be disconnected.  This leads us to an obvious
generalization of previous approaches to robustness, in which we focus on
the set or sets of vertices in a network that are connected by at least two
paths.  Such sets are called \defn{bicomponents}.  In this paper we study
the robustness of networks, including both real and model networks, in
terms of the size and behavior of their bicomponents.

Two paths connecting the same pair of vertices in a network are said to be
\defn{vertex-independent}, or simply \defn{independent}, if they share none
of the same vertices other than the starting and ending vertices.  A
$k$-component is a maximal subset of the vertices of a network such that
every vertex in the subset is connected to every other by $k$ independent
paths~\cite{West96}.  For the special cases of $k=2,3$ the $k$-components
are also called \defn{bicomponents} and \defn{tricomponents}, respectively.
Note that not all vertices need belong to a $k$-component for $k\ge2$,
which contrasts with the case for ordinary components ($k=1$), where every
vertex belongs to a component.

The vertices in a bicomponent have the property that no two can be
disconnected by the failure of any other single vertex.  Another
observation that will become important shortly is that the $k$-components
of a network are nested.  That is, every bicomponent is, trivially, a
subset of an ordinary component, every tricomponent is a subset of a
bicomponent, and so forth.  In this paper we concentrate primarily on
bicomponents, although we will give some results for $k\ge3$ where
appropriate.

To gain an understanding of the behavior of network bicomponents, let us
look first at a standard model network, the widely studied ``configuration
model,'' which is a network chosen uniformly at random from the set of all
networks with a given degree sequence~\cite{Luczak92,MR95,NSW01}.  The
probability of an edge falling between two vertices~$i$ and~$j$ in such a
network is $k_ik_j/2m$, where $k_i$ is the degree of vertex~$i$ and $m$ is
the total number of edges in the network.  The configuration model is a
useful guide to the expected qualitative behavior of many network
statistics, and in particular provides good insight into the behavior of
ordinary components (1-components), having at most one giant component of
size~$\Ord(n)$ and $\Ord(n)$ small components of size~$\Ord(1)$, a pattern
that is seen in most real-world networks as well.

A first interesting result to note is that, by contrast with the case for
1-components, there are in general (almost) no small bicomponents in the
configuration model.  To see this consider the small 1-components of the
configuration model, which, as shown elsewhere, are generally tree-like,
meaning they are not bicomponents.  In order to turn them into
bicomponents, we would need to add at least one edge to the tree, thereby
closing a loop and creating two paths between some vertices.  Since the
small components have size~$\Ord(1)$ and hence also $\Ord(1)$ pairs of
vertices, there are $\Ord(1)$ opportunities to perform such a closure in
each small component.  Each closure occurs with probability $k_ik_j/2m =
\Ord(n^{-1})$ on a sparse network and hence the total probability of
converting a small component into a bicomponent is~$\Ord(n^{-1})$.  Since
there are $\Ord(n)$ small components, this means that the total number of
small bicomponents formed in this way is~$\Ord(1)$.  Small bicomponents can
also be formed out of tree-like subsets of the giant component, but a
similar argument shows that such bicomponents are also $\Ord(1)$ in number.

Thus in the limit of large network size the probability that a randomly
chosen node belongs to a small bicomponent vanishes as $1/n$.  Speaking
loosely, there are no small bicomponents in the network.

There may, however, be a \defn{giant bicomponent}, and moreover it turns
out to be possible to calculate the expected size of the giant bicomponent
exactly in the limit of large graph size.  In order for a vertex to belong
to the giant bicomponent at least two of the edges incident on that vertex
must lead to the giant bicomponent by independent paths.  However, since
the giant bicomponent is a subset of the giant 1-component, it follows that
any edge that leads to the giant 1-component also leads to the giant
bicomponent, so it will suffice that at least two of our vertex's neighbors
be in the giant 1-component.

The probability that the neighbor of a vertex belongs to the giant
component is straightforward to calculate~\cite{NSW01}.  Let the degree
distribution of our network be~$p_k$, meaning that a randomly chosen vertex
has degree $k$ with probability~$p_k$.  If we choose an edge at random and
follow it to one of the vertices at its ends, then the number of edges
incident on that vertex, other than the one we arrived along, follows a
different distribution, the so-called \defn{excess degree distribution}:
\begin{equation}
q_k  = {(k+1)p_{k+1}\over\av{k}},
\end{equation}
as shown in~\cite{NSW01}, for example.  Here $\av{k}=\sum_k kp_k$ is the
mean degree for the entire network.

Now let $u$ be the probability that upon following an edge we reach a
vertex that does not belong to the giant component.  In order for this to
be the case, it must be that none of the other edges attached to that
vertex lead to vertices in the giant component, which happens with
probability~$u^k$, where $k$ is the excess degree.  Averaging over the
distribution~$q_k$ of~$k$, we then find that
\begin{equation}
u = \sum_{k=0}^\infty q_k u^k = G_1(u),
\label{eq:fixedpoint}
\end{equation}
where $G_1(z) = \sum_k q_k z^k$ is the probability generating function
for~$q_k$.  If the network is to have a giant component this equation must
have a nontrivial solution $u<1$.  It is straightforward to show that this
occurs if $G_1'(1)>1$, which leads to the well-known criterion of Molloy
and Reed~\cite{MR95} for the existence of the giant component.

Armed with these results we can now write down the probability that a
randomly chosen vertex belongs to the giant bicomponent.  The probability
that it does not is the probability that either zero or one, but not more,
of its edges lead to vertices in the giant component, which is
\begin{equation}
\sum_k p_k u^k + \sum_k k p_k (1-u) u^{k-1}
  = G_0(u) + (1-u) G_0'(u),
\end{equation}
where $G_0(z) = \sum_k p_k z^k$ is the probability generating function
for~$p_k$.  Then the probability~$S_2$ that the vertex \emph{is} in the
giant bicomponent is one minus this quantity:
\begin{equation}
S_2 = 1 - G_0(u) - (1-u) G_0'(u).
\label{eq:s2}
\end{equation}
Alternatively, $S_2$~is the expected size of the giant bicomponent as a
fraction of the size of the entire network.

We have not, in this derivation, demonstrated that the two paths from our
vertex to the giant bicomponent are independent.  However, since the
diameter of a random graph is~$\Ord(\ln n)$~\cite{BR04}, and since the
diameter provides an upper bound on the lengths of the paths in our
derivation, the probability of their intersecting is $\Ord(n^{-1} \ln^2
n)$, which vanishes as $n\to\infty$, so our result will be correct provided
that our network is large.

There can of course be no giant bicomponent when there is no giant
1-component, since the former is a subset of the latter.
Equation~\eqref{eq:s2} shows that, in general, the reverse is also true:
when there is a giant 1-component there is also a giant bicomponent.  Only
in the special case where $1-G_0(u)=(1-u)G_0'(u)$ can the giant bicomponent
vanish.

Thus the giant bicomponent appears, in general, at the same time as the
giant component: for any specific family of degree distributions, if there
is a transition marking the appearance of the giant component, the same
transition also marks the appearance of the giant bicomponent.

Equation~\eqref{eq:s2} can be generalized straightforwardly to
$k$-components for general~$k$.  The general result is
\begin{equation}
S_k = 1 - \sum_{m=0}^{k-1} {(1-u)^m\over m!}\,{\d^m G_0\over\d z^m}
                           \biggr|_{z=u},
\label{eq:sk}
\end{equation}
which implies that in fact $k$-components for \emph{all} $k$ appear at the
same time as the giant 1-component.

Since the presence of large $k$-components is, as we have said, a desirable
property for many kinds of networks, it is instructive to ask how robust
the property is to the removal of vertices or edges.  We already know, from
the results above, that if we start removing vertices from a network, the
giant $k$-component will vanish at the same time as the giant component
does.  Thus the thresholds for the complete destruction of robust
connectivity at all orders coincide.  This however does not tell the whole
story.  Using a simple variant of the arguments above we can also calculate
the size of the giant $k$-component as vertices are removed.

Let $r_k$ be the probability that a vertex of degree~$k$ is operational
(i.e.,~it hasn't failed or been removed from the network).  Common choices
for $r_k$ are $r_k=\mbox{constant}$, which corresponds to uniformly random
failure of vertices, or $r_k=\theta(k_\mathrm{max}-k)$, where $\theta(x)$
is the Heaviside step function, which corresponds to removal of all
vertices with degree $k>k_\mathrm{max}$, a form of targeted attack against
the best-connected vertices~\cite{AJB00}.

\begin{figure*}
\begin{center}
\includegraphics[width=10.5cm]{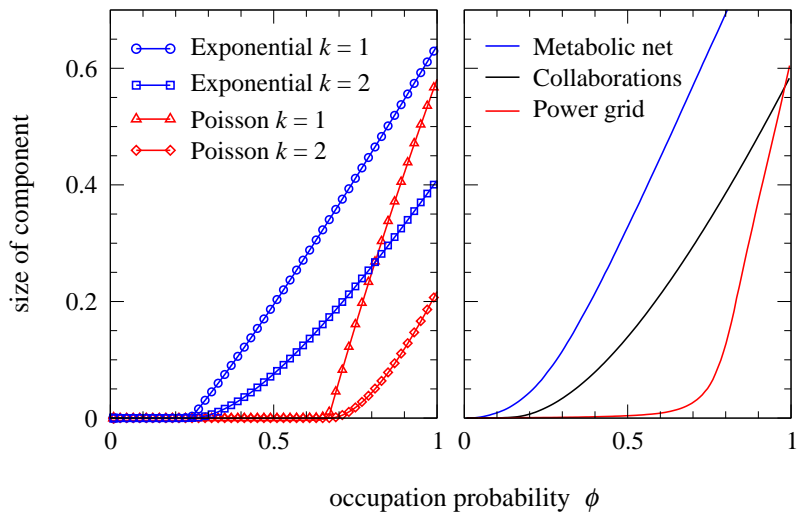}
\end{center}
\caption{Left panel: Size of giant component ($k=1$) and giant bicomponent
  ($k=2$) as vertices are randomly removed from random graphs with
  exponential ($\e^{-\lambda k}$ with $\lambda=0.4$) and Poisson (mean 1.5)
  degree distributions.  Solid lines are the analytic solutions; points are
  average numerical results for 100 simulations each on networks with
  $10^6$ vertices.  Error bars are smaller than the points in all cases.
  Right panel: Size of giant bicomponent as vertices are randomly removed
  from three real-world networks, a metabolic network for
  \textit{C.~Elegans}~\cite{DA05}, a collaboration network of scientists
  working in condensed matter physics~\cite{Newman01a}, and the Western
  States Power Grid of the United States~\cite{WS98}.}
\label{fig:results}
\end{figure*}

Then the probability~$u$ that an edge does not lead to a vertex in the
giant component, if that vertex has excess degree~$k$ (or total
degree~$k+1$), is the probability $1-r_{k+1}$ that the vertex has been
removed plus the probability $r_{k+1} u^k$ that it has not been removed but
that none of its other edges lead to the giant component either.  Averaging
over the distribution~$q_k$ of the excess degree, we then find
\begin{equation}
u = 1 - F_1(1) + F_1(u),
\label{eq:uf1}
\end{equation}
where $F_1(z) = \sum_k q_k r_{k+1} z^k$.  Then, by an argument similar to
the one leading to Eq.~\eqref{eq:sk}, the expected size of the giant
$k$-component is
\begin{equation}
S_k = F_0(1) - \sum_{m=0}^{k-1} {(1-u)^m\over m!}\,{\d^m F_0\over\d z^m}
                           \biggr|_{z=u},
\label{eq:skperc}
\end{equation}
where $F_0(z)=\sum_k p_k r_k z^k$.

The percolation transition at which the giant 1-component in a network is
destroyed is typically second-order in nature.  For $k$-components with
$k\ge2$ on the other hand, we can show using the results above that the
transition is of higher order.  Consider for instance the case of uniform
random failure of vertices, for which $r_k=\phi$ independent of~$k$.  Then
Eq.~\eqref{eq:uf1} becomes $u = 1 - \phi + \phi G_1(u)$ and, writing
$u=1-\epsilon$ and making use of $G_1(1)=1$, we find
\begin{equation}
\epsilon = \phi [ \epsilon G_1'(1) - \half \epsilon^2 G_1''(1) ] +
  \Ord(\epsilon^3)
\end{equation}
or, rearranging,
\begin{equation}
\epsilon = {2[\phi G_1'(1) - 1]\over\phi G_1''(1)}.
\label{eq:epsilon}
\end{equation}
In other words, $1-u$ is linear in $\phi-\phi_c$, where the critical
occupation probability is $\phi_c=1/G_1'(1)$~\cite{CEBH00}.

Now consider Eq.~\eqref{eq:skperc} for the size of the giant $k$-component
and let us rewrite it as follows.  Performing a Taylor expansion of
$F_0(z)$ about $z=u$ we find
\begin{equation}
F_0(z) = \sum_{m=0}^\infty {(z-u)^m\over m!}\,{\d^m F_0\over\d z^m}
                           \biggr|_{z=u}.
\end{equation}
Setting $z=1$ and making use of~\eqref{eq:skperc}, we then find that
\begin{equation}
S_k = \sum_{m=k}^\infty {(1-u)^m\over m!}\,{\d^m F_0\over\d z^m}
                           \biggr|_{z=u}.
\end{equation}
The leading order term in this expression is $\Ord(1-u)^k$ and hence, by
Eq.~\eqref{eq:epsilon}, $S_k = \Ord(\phi-\phi_c)^k$.

Thus the system displays a (potentially) infinite series of continuous
phase transitions marking the appearance of the giant components for
different values of~$k$, each occurring at the same point~$\phi_c$ but each
of different order, the transition for any given $k$ being of $(k+1)$th
order in~$\phi$.  This means that although, for example, the giant
bicomponent does appear at the same moment as the ordinary giant component,
its appearance is a third-order transition and hence it initially grows in
size at a far slower rate than the giant component so that, in practice,
the network does not achieve a significant level of robustness in the sense
considered here until considerably above~$\phi_c$.  This behavior is
illustrated in the left-hand panel of Fig.~\ref{fig:results}, where we show
the size of the largest 1- and 2-components in random graphs with two
different degree distributions, along with simulation results for the same
networks.  The second- and third-order natures of the phase transitions can
be clearly seen.

\begin{table}
\setlength{\tabcolsep}{4pt}
\begin{tabular}{l|llll}
\multicolumn{4}{c}{}                                & small \\
network            & $n$        & $S_1$   & $S_2$   & bicomp. \\
\hline
Internet (AS)      & $22\,963$  & $1$     & $0.651$ & $0.012$ \\
world wide web     & $325\,729$ & $1$     & $0.414$ & $0.076$ \\
power grid         & $4941$     & $1$     & $0.615$ & $0.062$ \\
\textit{C. Elegans} neural
                   & $297$      & $1$     & $0.949$ & $0$ \\
\textit{C. Elegans} metabolic
                   & $453$      & $1$     & $0.934$ & $0.049$ \\
physics collaborations   & $16\,726$  & $0.829$ & $0.588$ & $0.243$ \\
network scientists & $1589$     & $0.239$ & $0.084$ & $0.634$ \\
friendship         & $795$      & $0.979$ & $0.940$ & $0$ \\
dating             & $573$      & $0.503$ & $0.072$ & $0.014$
\end{tabular}
\caption{Statistics of a number of real-world networks.  The second to
  fifth columns give the number of vertices in the network, the fractions
  occupied by the largest component and bicomponent, and the fraction
  occupied by small components.  The networks are, in order, a snapshot of
  the Internet topology at the autonomous system (AS) level, the
  symmetrized web graph of a university web site~\cite{AJB99}, the Western
  United States power grid~\cite{WS98}, the neural~\cite{WS98} and
  metabolic~\cite{DA05} networks of the nematode \textit{C. Elegans},
  coauthorship networks of physicists~\cite{Newman01a} and network
  scientists~\cite{Newman06c}, and friendship and dating networks from a
  study of US school students~\cite{BMS02}.}
\label{tab:results}
\end{table}

Armed with the theoretical insights afforded by the configuration model,
let us turn now to the behavior of bicomponents in real-world networks.
All bicomponents in a network can be found in time $\Ord(n)$ using the
depth-first search based algorithm of Hopcroft and Tarjan~\cite{HT73a}.
Table~\ref{tab:results} summarizes the results of applying this algorithm
to a variety of previously documented networks.  The table reveals some
interesting features.  First we note that, with two exceptions, the
networks all have large giant bicomponents.  Certainly the giant
bicomponents are smaller than the giant components, but in each case the
networks have a substantial fraction of robust connections in the sense
considered here.  The two exceptions are the collaborations of network
scientists and the high-school dating network.  The former has quite a
small giant component, so the giant bicomponent cannot be very large,
although it is still quite a small fraction of the giant component size.
The dating network, however, is clearly anomalous, having a very
substantial giant component but a very small giant bicomponent.  It is
interesting to consider whether there might be sociological reasons for
this anomaly.

Second, we note that in all but two cases the networks have no small
bicomponents, or very nearly none.  This observation agrees well with our
calculations for random graphs above, but is otherwise somewhat surprising.
It has been observed that most real-world networks contain a high density
of short loops~\cite{WS98}, a feature that random graphs lack and one that
might be expected to give rise to small bicomponents.  Our observations
imply, however, that in most cases the loops are attached to the giant
bicomponent, rather than forming independent bicomponents, and that most
portions of the network not in the giant bicomponent are tree-like.  This
agreement with the random graph model stands in sharp contrast with studies
of other network properties such as clustering and assortative mixing, in
which random graphs match real networks poorly.  Note also the exception to
the overall pattern provided by the two collaboration networks: both have
quite significant fractions of their vertices in small bicomponents, a
feature possibly rooted in the particular social structure of scientific
research.

It is also interesting to examine the robustness of the bicomponent
structure to removal of vertices---either random or targeted---as we did
for our model networks.  The Hopcroft--Tarjan algorithm is a poor choice
for this calculation, since we would have to perform $n$ runs of the
algorithm to find the bicomponents after the removal of each vertex, for a
total running time~$\Ord(n^2)$.  For the larger networks studied this is
prohibitive, so instead we have developed a different algorithm that allows
the calculation to be performed much faster.  The algorithm is similar in
spirit to the fast percolation algorithm of Newman and Ziff~\cite{NZ00} and
will be discussed in detail elsewhere.  Here we give only a brief
description of its working.

The algorithm starts with an empty network and adds vertices, rather than
taking them away, and avoids finding the bicomponents anew after each
addition by calculating only the change in the bicomponent structure from
the previous step, which is usually minor.  The algorithm stores the
structure of the components and bicomponents in two separate ``forest''
data structures (i.e.,~sets of trees) with one tree for each (bi)component.
Vertices within components contain pointers that point to others in the
same component, such that by following a sequence of such pointers we can
reach the root of the tree, thereby identifying the component uniquely.  As
each new vertex is added to the network we check in this way to which
components its neighbors belong, amalgamating those components if necessary
by adding a pointer from the root of one tree to the root of the other.  If
the added vertex joins neighbors that already belong to the same component,
a loop and hence new bicomponent has been created, or old bicomponents
extended or joined, and the bicomponent trees are updated appropriately.
The tree traversals employed by the algorithm take $\Ord(\ln n)$ time on
average and hence the algorithm can add all $n$ vertices in a total running
time~$\Ord(n\ln n)$.  In practice, this gives an improvement in running
time of a factor 100 or more over the Hopcroft--Tarjan algorithm for the
networks studied here, and renders the calculations easily doable on a
standard desktop computer.

The right-hand panel of Fig.~\ref{fig:results} shows the results of the
application of this algorithm to three of the networks from
Table~\ref{tab:results}, the metabolic network, the physicist
collaborations, and the power grid.  For each network there appears to be a
transition point below which the giant bicomponent is destroyed and the
network can no longer be said to be robustly connected.  For two of the
networks, however, the transition appears to be at or close to $\phi=0$,
indicating that the networks are highly robust in the sense considered
here: nearly all the vertices have to be removed from the network before
the giant bicomponent is destroyed.  The third network, the power grid,
shows a much higher transition probability, indicating that this network is
relatively fragile to random node removal.  On the other hand, we also see
in all cases that the transition at which the giant bicomponent appears is
a gradual one; the gradient at the transition is shallow---perhaps even
zero---so that the giant bicomponent grows very slowly at first above the
transition.

Taken together, the analytic and numerical results presented here give an
interesting picture of the behavior of network bicomponents.  Real-world
networks appear to be quite robust in the sense of having large giant
bicomponents and moreover the existence of these bicomponents is in some
cases (though not all) itself robust to the deletion of vertices.  In
practice, however, although the giant bicomponent may persist as vertices
are removed from the network, its size dwindles rapidly so that large
portions of the network lose robust connection considerably before the
transition point at which the giant bicomponent finally vanishes.  In each
of these respects the behavior of our networks is surprisingly similar to
the behavior of the exactly solvable configuration model, which predicts a
giant bicomponent that persists down to the point at which the ordinary
giant component disappears, but with an unusual third-order transition at
that point that ensures that the size of the bicomponent will be small as
we approach the transition.

The authors thank Cris Moore for useful conversations.  This work was
funded in part by the National Science Foundation under grant DMS--0405348
and by the James S. McDonnell Foundation.

\end{document}